\documentclass[epj,a4paper]{svjour}
\usepackage{graphics}
\usepackage{psfig}
\begin{document}

\newcommand{\beq}{\begin{equation}}
\newcommand{\eeq}{\end{equation}}
\newcommand{\sgn}{{\rm sgn}}
\newcommand{\beqa}{\begin{eqnarray}}
\newcommand{\eeqa}{\end{eqnarray}}
\newcommand{\refe}[1]{(\ref{#1})} 
\newcommand{\text}[1]{\rm #1}

\newcommand{\rem}[1]{}

\title{Steps and facets at the surface of soft crystals}

\author{
P.~Nozi\`eres\inst{1} 
\and
F.~Pistolesi\inst{2}
\and
S.~Balibar\inst{3}
}
\institute{
Laboratoire d'Etude des Propri\'et\'es Electroniques des Solides
CNRS, B.P. 166, 38042 Grenoble Cedex 9, France
\and
European Synchrotron Radiation Facility,
B.P. 220, 38043 Grenoble Cedex 9, France
\and
Laboratoire de Physique Statistique de l'Ecole Normale Sup\'erieure 
associ\'e aux Universit\'es Paris 6 et Paris 7 et au CNRS\\
 24 rue Lhomond, 75231 Paris Cedex 05, France
}
\date{Received: 26 June 2001}
\abstract{
We consider the shape of crystals which are soft in the sense that
their elastic modulus $\mu$ is small compared to their surface tension
$\gamma$, more precisely $ \mu a < \gamma$ where $a$ is the lattice
spacing. 
We show that their surface steps penetrate inside the crystal as edge
dislocations. 
As a consequence, these steps are broad with a small energy which we 
calculate.
We also calculate the elastic interaction between steps a distance $d$ 
apart, which is a $1/d^2$ repulsion.
We finally calculate the roughening temperatures of
successive facets in order to compare with the remarkable shapes of
lyotropic crystals recently observed by P. Pieranski {\em et
al}\cite{Pieranski,EPJ}. Good agreement is found.
}
\maketitle

\section{Introduction}
\label{secI}

An elastic material is characterized by an elastic modulus $\mu $ which is
an energy per unit volume. Any interface, whether with a vapour or a
liquid phase,
has a surface energy $\gamma$ per unit area. The ratio $\gamma /\mu
$ is thus a characteristic length which should be compared to other lengths
in the problem, for instance the lattice spacing $a$. More specifically, let
us define
\begin{equation}
	d_{0}
	=
	\frac{\left( 1-\sigma \right) \gamma }{\mu }
	\label{eq:d0}
\end{equation}
where $\sigma $ is Poisson's modulus. Three situations may occur:

\noindent (i) For usual materials in contact with their vapour, 
$d_{0}$ is smaller than $a$. Indeed, according to Frank and
Stroh\cite{Frank}, a usual value for $\gamma / \mu$
is $ 0.1 a $ and Poisson's modulus is usually equal to 1/3. As a
consequence, we will see that steps are edge dislocations bound to the 
surface. The familiar
``terrace, ledge, kink'' (``TLK'') model\cite{TLK,Villain} holds. 
\noindent (ii) At a solid-melt interface $\gamma $ is smaller
 because the solid density is not very different from that of the liquid. 
Then $d_{0}\ll a$: elastic strains due to surface 
pertubations like steps are only small corrections.

\noindent (iii) It may happen that $d_{o}$ is larger than $a$ : then
it pays to spend elastic energy in order to gain surface energy. As we 
shall see below, steps are dislocations which are buried a depth h below 
the surface. A new length scale appears in the problem. The steps are broad 
and their energy is strongly reduced by an elastic relaxation. That
should always occur if the material is soft enough. A spectacular
illustration of this situation is given by the lyotropic crystals
recently studied by Pieranski et al. \cite{Pieranski,EPJ}. These
 crystals are water solutions of non-ionic surfactant molecules which
arrange in a loose cubic structure with $a$ of order 50 $\mathrm\AA$.  The
structure is soft ($\mu = 10^7 \mathrm erg/cm^3$), but the surface 
tension is only a
little less than that of pure water ($\gamma = 25\: \mathrm dyn/cm$). 
As a result,
the ratio $d_0/a$ is much larger than in usual crystals.
 This is a
typical
example of the anomalous case of ``soft crystals'' which we consider 
in this article\cite{remark}.

Pieranski et al.\cite{Pieranski,EPJ} have observed that the facetting
of their lyotropic crystals is remarkable: up to 60 different facets
are present on the equilibrium shape. Our first motivation was to
understand the origin of the large number of facets on these
crystals. As we shall see, it is the large size of the lattice spacing
$a$. Pieranski et al. also invoked the large value of 
the step-step repulsion, but, as we shall see, it is compensated by the 
small value of the step energy in the calculation of the roughening 
temperatures. In section II, we calculate the unusual structure and
 the energy of steps in such soft crystals.
In section III, we calculate the interaction energy of two steps a distance 
$d$ apart, and we show that it is a $1/d^2$ repulsion whose origin is 
surface capillary forces within the step profile.
In Section \ref{secIV} we calculate the net macroscopic surface energy of a
vicinal surface, with regularly spaced steps, which is the starting
point for a discussion of shapes. Section \ref{secV} is devoted to the
roughening transition of such vicinal surfaces. In the last section,
we use our results to calculate how many facets should be observed on the 
crystals of Pieranski et al.\cite{Pieranski,EPJ} and we compare with their 
experimental results.

\section{Structure and energy of a single step}
\label{secII}
For simplicity we consider a simple cubic crystal with lattice spacing
$a$.  If no relaxation occurred, a single step on a (100) facet would
expose an extra surface $a$ (per unit length of the step). 
In this case, the step energy $\beta$ would be $\gamma a$. 
Such a step is a remnant of an edge dislocation along the $y$ axis with
a Burgers vector along the $z$ axis as it moves towards the
surface.
One way to reduce its energy is to move it inside the crystal as an
edge dislocation which is now a depth $h$ below the free surface
$z=0$. A similar phenomenon was considered by Lejcek et al.\cite{Lejcek} 
in the 
different context of smectic A liquid crystals.
 The dislocation introduces an extra lattice
plane on half the area and produces a rounded step at the
surface. Burying the dislocation rounds off the step and decreases the
surface energy. The price to pay is the dislocation elastic energy. As
we shall see, this price is low for soft crystals, but it would be too
high for usual crystals.

In an infinite medium, the displacement $u_{i}^{(o)}$ and stress
 $\sigma _{ij}^{(o)}$ created by
the edge dislocation are known \cite{Landau}, 
as well as the total elastic energy per unit length
\beq
	\beta^{(o)}
	=
	\frac{\mu a^{2}}
	{4\pi \left( 1-\sigma \right) }
	\ln \frac{R_{\max }}{r_{o}}
\eeq
where $r_{o}$ is an appropriately defined core radius of the order of the 
lattice spacing $a$. 
$R_{\max }$ is the size of the system which will be replaced by $h$ in
the presence of a free surface (images suppress divergences at
infinity). Qualitatively we can guess the final result easily: the
step width at the surface is of order $h$ and the profile lengthening
is thus $\approx a^{2}/h$. 
Balancing $\beta^{(o)}$ against the capillary energy $\gamma a^{2}/h$
yields at once 
$ h \approx \gamma ( 1-\sigma ) /\mu = d_{o}$ :
steps are buried in soft crystals!
 But a quantitative discussion requires more
effort! 
In order to calculate this elasticity problem, we shall proceed in four 
successive steps. As a zeroth 
order approximation, we will start with an infinite medium, and 
our notations for  
stresses and displacements will have a superscript (o). 
Suppressing the upper half space 
modifies the strain field in the lower part, thereby adding an 
extra displacement $u^{(1)}$ at the surface $z=0$.
The energy stored below is modified, and we must subtract the 
energy originally stored above. In a final step we must add
a correction $u^{cap}$ due to the existence of surface tension.

Since we first ignore the effect of surface tension, the normal stresses
$\sigma _{zz}^{(o)}$ and $\sigma _{zx}^{(o)}$ have to be cancelled at the 
crystal surface. 
If we dealt with screw dislocations, we would treat the elasticity of 
a half space by introducing image dislocations. However, this method does not
work with edge dislocations: adding a symmetric edge dislocation
 with opposite Burgers
vector does cancel out the stress $\sigma _{zz}$, but it doubles $
\sigma _{zx}$! We are thus forced to solve the elastic problem
directly. In an infinite medium, the upper half exerts a force
 $ F_{i}^{(o)}=\sigma _{iz}^{(o)}$ on the lower half. 
Cancelling this force is the same as applying an opposite force
$-F_{i}^{(o)}$, resulting in an extra displacement $u_{i}^{(1)}$ and
stress $\sigma _{ij}^{(1)}$  on top of the original
$u_{i}^{(o)},$ $\sigma _{ij}^{(o)}$ . 
This extra displacement \emph{at the surface} is
\beq
	u_{i}^{(1)}\left( x\right) 
	=
	\int_{-\infty }^{+\infty }\chi _{ij}\left(x-x^{\prime }\right) 
	\,\left( -F_{j}
	\left( x^{\prime }\right) \right)\,dx^{\prime }
\eeq
where $\chi _{ij}\left( x-x^{\prime }\right)$ is the two dimensional
surface response function, given for instance in 
Landau-Lifschitz \cite{Landau} (see eq. A6 in the appendix). 
The net normal elastic displacement that controls the surface profile
is
\beq
	u_{z}^{el}=u_{z}^{(o)}+u_{z}^{(1)}
\eeq
In order to calculate the elastic energy, two corrections are needed

\noindent 
({\em i}) 
Add the energy lowering upon relaxation of the forces 
$F_{i}^{(o)}$
\beqa
	\lefteqn{\delta \beta^{(1)} =
	\int_{-\infty }^{+\infty }dx\int_{0}^{1}u_{i}^{(1)}d\lambda
	\sigma _{zi}^{(o)}\left( 1-\lambda \right) }
	&& \nonumber \\
	&=&
	-\frac{1}{2}\int_{-\infty }^{+\infty }dx\,dx'
	\chi_{ij}\left( x-x^{\prime }\right) 
	\sigma_{zi}^{(o)}
	\left( x\right) 
	\sigma _{zj}^{(o)}
	\left( x^{\prime }\right)
\eeqa
({\em ii}) 
Subtract the part of $\beta^{(o)}$ that was stored above the surface. The
bottom exerts on the top a force $-F_{i}^{(o)},$ which by itself would
create a displacement
\beq
	u_{i}^{(2)}\left( x\right) 
	=
	\int_{-\infty }^{+\infty }
	\overline{\chi }_{ij}
	\left( x-x^{\prime }\right) 
	\,\left( -F_{j}^{(o)}
	\left( x^{\prime}\right) \right) 
	\,dx^{\prime }
\eeq
This displacement is calculated in a way similar to the previous one 
($u^{(1)}$). As explained by Landau\cite{Landau}, it involves a response 
function $\overline{\chi }_{ij}$ which is the transposed of $\chi _{ij}$, because
the medium now sits above instead of below. 
One has to substract an energy
\beq
	\delta \beta^{(2)}
	= 
	-\frac{1}{2}\int_{-\infty }^{+\infty }dx\,u_{i}^{(2)}
	\left(x\right) 
	\left[ -F_{i}^{(o)}\left( x^{{}}\right) \right]
\eeq
The cross terms cancel out between $\chi $ and $\overline{\chi }$ and
we are left with a simple result for the elastic energy at this stage:
\beq
	\beta^{\rm el} 
 	=
	\beta^{(o)}-\,\int_{-\infty }^{+\infty }\!\!\!\!\!
	dx\,dx'
	\sum_{i=x,z}
	\left[	\chi _{ii} \left( x-x'\right) 
		\sigma _{zi}^{(o)}\left( x \right)
		\sigma _{zi}^{(o)}\left( x^{\prime }\right) 
	\right]
	\,.
\eeq

As explained in the appendix, we know $\chi _{ij,}$ so that we can solve
 this elastic problem
analytically. Note that we could also use 
 the method of images if we added counterforces $2F_{x}^{(o)}$ 
along the surface only.
Comparing the two approaches is a useful check of a rather cumbersome
algebra.

The calculation is most easily carried out in Fourier space, according to
the definitions
\beq	
	\left\{
	\begin{array}{rcl}
	f\left( x\right) 
	&=&
	\displaystyle
	\int_{-\infty }^{+\infty } 
	f\left( k\right) e^{ikx} 
	dk
	\\
	f\left( k\right) 
	&=& 
	\displaystyle
	\frac{1}{2\pi }\int_{-\infty }^{+\infty } 
	f\left( x\right) 
	e^{-ikx}dx
	\end{array}
	\right.
	\,.
\eeq
For a periodic force $u_{i}\left( k\right) =\chi _{ij}\left( k\right)
F_{j}\left( k\right)$. It is easily verified that
\beq
	\chi _{ij}\left( x\right) 
	= 
	\frac{1}{2\pi }\int_{-\infty }^{+\infty }
	\chi_{ij}
	\left( k\right) e^{ikx} dk
\eeq
The elastic energy is
\beqa
	\lefteqn{
	\beta^{\rm el} = 
	\beta^{(o)}}
	&&\nonumber \\
	&-&\,2\pi \int_{-\infty }^{+\infty }dk\,\,
	\left[ 
	\left| \chi_{zz}
	\left( k\right) \sigma _{zz}^{(o)}\left( k\right) 
	\right| ^{2}
	+\chi_{xx}
	\left( k\right) \left| \sigma _{zx}^{(o)}\left( k\right) \right|^{2}
	\right]
	\nonumber \\ 
\eeqa
We only need  carry integrals!

But this is not the end of the story. We have not yet included elastic
deformations due to the existence of the surface tension, that is to 
capillarity!
 The normal stress $\sigma _{zz}$ at the
surface is not zero. The surface is under mechanical equilibrium because 
the surface stress is compensated by
 the Laplace force $\Phi _{z}$ which is exerted by the
surface on the inside. It is  a consequence of the curvature 
in the step profile:
\beq
\Phi _{z}=\gamma \frac{d^{2}u_{z}}{dx^{2}}
\eeq
That force creates an extra displacement $u^{\rm cap}$ which in turn
modifies the profile. We thus write in reciprocal space
\begin{eqnarray*}
	\Phi _{z}\left( k\right)  
	&=& 
	-\gamma k^{2} 
	\left(u_{z}^{ \rm el}+u_{z}^{ \rm cap}\right)  
	\\
	u_{z}^{\rm cap} \left( k\right)  
	&=& 
	\chi _{zz}\left( k\right) \Phi _{z}\left(k\right)
\end{eqnarray*}
whose solution is
\begin{eqnarray*}
	u_{z}^{ \rm cap}\left( k\right)  
	&=& 
	-\,\frac{\alpha _{k}}{1+\alpha _{k}} \,\,u_{z}^{\rm el} \\
	\alpha _{k} 
	&=& 
	\gamma k^{2}\chi _{zz}\left( k\right)
\end{eqnarray*}
Because the displacement $u^{\rm cap}$ modifies $\Phi
_{z}\left(k\right)$, the usual counterforce argument does not apply
and it is simpler to write the net capillary energy as the sum of its
bare value and the elastic energy stored in producing $u^{\rm cap}$
\beq
	\beta^{\rm cap} 
	=
	2\pi \int_{-\infty }^{+\infty }\!\!\!\!\!\!\!dk
	\left\{ 
	\frac{\gamma k^{2}}{2}
	\left| 
		u_{z}^{\rm el}\left( k\right) +u_{z}^{\rm cap}(k)
	\right|^{2}
	+\frac{\left| u_{z}^{\rm cap}\left( k\right) \right| ^{2}}
	{2\chi _{zz}
	\left( k \right) }
	\right\}
\eeq
(minimizing with respect to $u^{\rm cap}$ reproduces the expression of
$u^{\rm cap}$).  We thus find
\beq
	\beta^{\rm cap} 
	= 
	2\pi \int_{-\infty }^{+\infty }dk
	\frac{\gamma k^{2}}{2\left(1+\alpha _{k}\right) }
	\left| u_{z}^{\rm el} \right|^{2}\,
\eeq
As expected, the elastic relaxation reduces the original capillary energy.

Our program is now clear: first we calculate 
$\sigma ^{(o)} \left( k\right) $ and $\chi \left( k\right)$, 
then we minimize $\beta=\beta^{\rm el}+\beta^{\rm cap}$ 
with respect to the depth $h$. 

Finally we calculate the step energy, which is the total energy of the
dislocation. The algebra is done in the Appendix. It leads to
\beqa
	\lefteqn{\beta 
	=
	\frac{\mu a^{2}}{4\pi \left( 1-\sigma \right) }
	\int_{0}^{1/r_{o}}\frac{dk}{k}
	\left\{ 1
	\vphantom{\frac{\left( 1+kh\right)^{2}}{1+kd_{o}}}
	\right.
	}&& \nonumber \\
	&-&
	\left.
	e^{-2kh}
	\left[ 1+2kh+2k^{2}h^{2}-2kd_{o}
	\frac{\left( 1+kh\right)^{2}}{1+kd_{o}}
	\right] 
	\right\}
\eeqa
where $d_{o}=\left( 1-\sigma \right) \gamma /\mu $ is our
characteristic length. 
The term $1$ is the logarithm in $\beta^{(o)}$, the first three terms
in the square bracket are the contribution of counterforces and the
last one the effect of capillary forces. 
The denominator $\left( 1+kd_{o}\right) $ reflects the self consistent
calculation of $u^{\rm cap}$.
Note that capillarity is hidden in the single length $d_{o}$. We
rewrite the curly bracket as
\beq
	1-e^{-2kh}-
	2kh\, e^{-2kh}\left[ 1-\frac{d_{o}}{h}\right] 
	\left[ 1+kh\frac{1-d_{o}/h}{1+kd_{o}}\right]
\eeq
It is easily verified that this quantity is stationary with respect to $h$
when $h=d_{o}$.
That result is an exact version of our original estimate.  Note that
minimization is the same {\em for every} $k$: as a consequence it
will persist for two steps or for a regular array. Altogether the
total energy takes an incredibly simple form
\beqa
	\beta
	&=&
	\frac{\mu a^{2}}{4\pi \left( 1-\sigma \right) }
	\int_{0}^{1/r_{o}}\frac{dk}{k}
	\left[ 1-e^{-2kd_o} \right]
	\label{beta1}
	\\
	&=&
	\frac{\mu a^{2}}{4\pi \left( 1-\sigma \right) }
	\left[\Gamma+\ln \left({2d_o\over r_o}\right)\right]
\label{eq:beta}
\eeqa
where $\Gamma\approx 0.577 $ is Euler's constant.
All complications have disappeared! In view of the messy intermediate
algebra there is probably a reason for that, but we do not know it.

Eq. \refe{eq:beta} is our first conclusion and it is non trivial:
 it shows that the energy
scale for steps is not $\gamma\, a$, but $\mu\, a^2$, down 
by a factor $\mu \, a /\gamma \approx a/d_o$. This  
will affect the roughening transition temperature of vicinal surfaces 
drastically. Step burying is physically important.

\section{The interaction between steps}
\label{secIII}

The interest of Fourier space calculations is the fact that they are 
immediately extendable to several steps. 
For a pair of steps a distance $d$ apart, the strain and stress 
tensors are simply multiplied by a factor $(1+e^{ikd})$.
The resulting energy integral for two steps is modified by a factor:
\beq
	 |1+e^{ikd}|^2 = 2 \left(1+\cos kd\right)
\,.
\eeq
Note that the depth $h=d_o$ is unchanged, as minimization was
achieved for {\em each} $k$.
The interaction between steps is contained in the term 
$2 \cos kd$. Using the above result for the energy of an isolated 
step we find that the energy for the two steps becomes:
\beqa
	2\beta(d)
	&=&
	\frac{\mu a^{2}}{2\pi \left( 1-\sigma \right) }
	\int_{0}^{1/r_{o}}\frac{dk}{k}
	\left[ 1-e^{-2kd_o} \right]
	\left[ 1+\cos kd \right]
	\nonumber
	\\
	&=&
 	 \frac{\mu a^{2}}{2\pi \left( 1-\sigma \right) }
	\left[
	\Gamma + \ln \left({2d_o \over r_o}\right) + 	 	
	{1 \over 2} 
	\ln
	\left(
	1 +{4 d_o^2 \over d^2 }
	\right)	
	\right]	\,.
	\nonumber\\
	\label{beta2}
\eeqa
The interaction energy is contained in the last term and for large 
$d$ it is:
\beq
	\epsilon_{12}=
	 \frac{\mu a^{2} d_o^2 }{\pi \left( 1-\sigma \right) d^2 }
	\label{interaction}
	\,.	
\eeq

It is interesting to compare this global repulsion with 
the successive results one would get at various steps of an 
analysis using image dislocations.

\noindent ({\em i}) Using only image dislocation without compensating
forces $2F^{(0)}$ leads to the interaction energy of four dislocations 
arranged in a rectangle with height $2h$ and base $d$.
We want half the total elastic energy $\epsilon$ of these four
dislocations since we integrate the elastic energy over a half
space. The starting point is the interaction energy of a pair of
parallel edge dislocations, with Burgers vectors $b_{1}$ and $b_{2}$
pointing in the same direction, but either parallel or
antiparallel. That interaction is given in the literature
\cite{Friedel,Hirth}
\begin{equation}
	\epsilon_{12} 
	= 
	\frac{\mu b_{1}b_{2}} 
	{2\pi \left( 1-\sigma \right) }
	\left[ 
		\ln 
		\frac{R_{\max }}{R}-\sin ^{2}\theta 
	\right]  
	\label{eq:epsilon12}
\end{equation}
where $R$ and $\theta $ are the polar coordinates of $2$ with respect
to $1$ (polar axis along $b$), and $R_{\max }$ an upper cutoff which
will disappear for dislocation dipoles. The first term in the bracket
yields a radial force which is repulsive for $b_{1}=b_{2}$ but
attractive if $b_{1}=-\:b_{2}$. The dislocation 1 is repelled by its
neighbour (dislocation 2), but it is attracted by the image 2' of this
neighbour. We also recover the known result that a double dislocation
with Burgers' vector $2b$ tends to split into two single dislocations
$b$. The second term in the bracket of Eq. \refe{eq:epsilon12} yields a
torque. For our rectangular configuration, and burgers vectors of 
amplitude $a$, we find
\beqa
	\lefteqn{
	\epsilon
	\: = \:
	\frac{1}{2}\:\frac{\mu a^{2}}{4\pi ( 1-\sigma )}
	\left[
		4\ln \frac{R_{max}}{r_0}-
		4\ln \frac{R_{max}}{2h}+
	\right.
	}\nonumber \\
	&&
	\left.
		4\left(\ln \frac{R_{max}}{d}-1\right)
		-4\left(\ln \frac{R_{max}\cos \alpha }{d}-
		\cos ^{2}\alpha \right)
	\right]
	\nonumber 
\eeqa
In the bracket the first term is the energy of individual dislocations, the
second the interaction of genuine dislocations $1$ and $2$ with their
respective images 1' and 2', the third $(1,2)$ and $(1',2')$,
the last one $(1,2')$ and $(1',2)$. The angle $\alpha $ is the tilt of cross
links, tan$(\alpha) =2h/d.$
That result holds for any value of $h/d$ as long as $h$ and $d$ are much
larger than $a$.
The elastic interaction between the steps corresponds to the last two terms
in the bracket
\begin{equation}
	\epsilon_i 
	=
	-\frac{\mu a^{2}}{2\pi \left( 1-\sigma \right) }\left[ 
	\sin ^{2}\alpha
	+\ln (\cos \alpha) \right]  
	\label{eq:Ud12i}
\end{equation}
For small $\alpha $, {\em i.e.} large distance $d$, the bracket is
$\alpha ^{2}/2$: the interaction is \textit{attractive}, proportional
to $1/d^{2}$ as befits a step interaction. It turns into a repulsion
for larger $\alpha$, the change corresponding to the minimum of
$\epsilon_i$ at $\alpha =\pi /4$, i.e. $d=2h$.
If true, such a behavior would lead to the prediction of
a sharp angular matching at the edge of a flat facet\cite{Beg-Rohu}. 

\noindent({\em ii})
The next approximation is the inclusion of counterforces at the surface.
 One finds that the $1/d^2$ attraction 
is {\em exactly} cancelled out. The net step interaction
is controlled by the next order in the expansion:
it is a $1/d^4$ {\em attraction}.

\noindent({\em iii}) 
But capillarity ruins the above result. As any relaxation mechanism it 
induces a $1/d^2$ {\em repulsion} that supersedes the 
$1/d^4$ term. We conclude that step interaction is  
controlled by the elastic strain field which is produced by the surface 
tension.
Just as in the classic picture by Marchenko and Parshin\cite{Marchenko}
 it is repulsive and $\sim 1/d^2$.
Our picture does not depart from standard wisdom. Interestingly, 
Eq. \refe{interaction} can be written as
\beq
\epsilon_{12}=
\frac{(1 - \sigma)\gamma ^2 a^2}{\pi \mu d^2 }
\eeq
which looks like Marchenko's result with a force doublet $\gamma a$ on the step; 
however, the force doublet is $\mu a^2$ distributed on a length $d_0$.

\section{The energy of a vicinal surface}
\label{secIV}

In practice we are not interested in a single pair of steps, but in a
regular array with period $d$, describing a vicinal surface with average
tilt $\overline{\phi}$ such that tan$\overline{\phi}=a/d$. Due to that tilt
the Burgers vector of dislocations is not exactly perpendicular to the real
surface, which strictly speaking makes the elastic calculation far more
complicated. We can ignore such a complication if we assume again that the
lengths $h$ and $d$ are both much larger than the atomic spacing $a$: 
the angle $\overline{\phi }$ is then small and we retain only the lowest order
in $a/d$ and $a/h$ while making no assumption on $h/d$. 
Once more, the depth $h$ is the same as for a single step.

We want to calculate the total energy per step $\beta$ and per 
unit area $U=\beta/d$. 
We can proceed in the same way we did for two dislocations.
Given the Fourier transform of the stress tensor of 
a single step $\sigma^{(o)}(k)$, we can generate the 
periodic superposition as follows:
\beq
	\sigma(k) = 
	\sum_{n=-N}^{+N} 
	e^{i\, k d n} 
	\sigma^{(o)}(k)
	\,,
\eeq
where $2N+1$ is the number of steps. 
If the total energy of a single step is 
\beq	
	\beta_1 = \int_{-\infty}^{+\infty} b(k) dk
\eeq
the total energy per step in a periodic arrangement becomes:
\beqa
	\beta
	&=& 
	{1\over 2N+1}
	\sum_{n,m=-N}^{+N} 
	\int dk \, 
	e^{i\, k d (n-m)} 
	b(k)
	\nonumber 
	\\
	&=&
	{2 \pi \over d }\sum_{\nu=-\infty}^{+\infty} 
	b\left({ 2\pi \nu  \over d} \right)
	\,.
\eeqa
Using the expression for $b(k)$ given in Eq.~\refe{beta1} we obtain:
\beq	
	\beta(d)
	= 
	{\mu a^2 \over 4 \pi (1-\sigma)}
	\left[
		{2 \pi d_o \over  d} + 
		\sum_{\nu=1}^{d/(2\pi r_o)} 
		{(1-e^{-4 \pi \nu d_o/ d}) \over \nu}
	\right]
	\,.
\eeq
The energy per unit area is $U=\beta(d)/d$.  The single term $\nu=0$
yields the capillary energy $\gamma a^2/2d^2$ associated to the
surface lenghtening due to an average tilt $\overline \phi$.
 
The summation can be done analytically for $d_o/r_o$ large and 
it gives:
\beqa
	\lefteqn{
	\beta(d)
	= 
	{\mu a^2 \over 4 \pi (1-\sigma)}
	\times
	}\nonumber \\
	&&
	\left[
	{2 \pi d_o\over  d}
	+\Gamma
	+\ln\left({d \over  2 \pi r_o  }\right)
	+
	\ln\left({1-e^{-4 \pi d_o/d}}\right)
	\right]\,,
	\nonumber
\eeqa
where we used 
$
	\sum_{\nu=1}^{\nu_m}  {1/\nu}
	=
	 \Gamma+\ln \nu_m + O(1/\nu_m)
$
and 
$
	\sum_{\nu=1}^\infty {e^{-\nu \lambda} /\nu}
	=
	-\ln(1-e^{-\lambda})
$.
We rearrange the expression in order to separate the core dependence.
We finally find:
\beqa
	\lefteqn{
	\beta(d)
	= 
	{\mu a^2 \over 4 \pi (1-\sigma)}
	\times
	}\nonumber \\
	&&
	\left[
	\Gamma
	+\ln\left({2 d_o\over  r_o }\right)
	+{2 \pi d_o\over  d}
	+
	\ln\left({1-e^{-4 \pi d_o/d} \over 4 \pi d_o/d}\right)
	\right]
	\label{betalattice}
	\,.
\eeqa
As mentioned above the relevant scale for the $d$ dependence 
of $\beta$ is $d_o$. 
The expansion of $\beta(d)$ for large $d$ begins with a 
positive $1/d^2$ term, indicating repulsion of steps:
\beq
	\beta(d) = 
	{\mu a^2 \over 4 \pi (1-\sigma)}
	\left[
	\Gamma
	+\ln\left({2 d_o\over  r_o }\right)
	+{2\pi^2 d_o^2 \over 3 d^2}
	\right]
	\label{betaexpansion}
	\,.
\eeq
The step-step interaction coefficient
is correctly  $\sum_{n=1}^\infty 1/n^2$ $=\pi^2/6$ times the 
two-step interaction energy given in \refe{interaction}.
For small $d$ the leading term is simply the capillary energy
needed to lengthen the surface in order to tilt it $\sim 1/d$.

\section{Roughening temperatures}
\label{secV}

It is now well known\cite{Beg-Rohu,Balibar94} that, as their
temperature decreases, crystals are facetted in an increasing number
of different directions.
In any rational direction, there is a roughening transition at a
temperature $T_R$. Below $ T_R$, the surface is ``smooth",
{\em i.e.} facetted or flat. It is localized by the influence of the crystal
lattice. Above $T_R$, the surface is ``rough" like a liquid
surface and rounded on the equilibrium shape of the crystal.  
The roughening temperature is given by the universal
relation\cite{Fisher,Balibar94}
\begin{equation}
	k_BT_R = \frac{2}{\pi}\, \gamma\, a^2
	\label{eq:univ}
\end{equation}
where  $a$ the periodicity of the surface energy along the normal to the 
interface, 
in practice the step height.
As for $\gamma$, it is the value of the surface stiffness at $T_R$. In
the case of interest here, where the lattice periodic potential is a
small perturbation (``weak coupling''), one can neglect the critical
variation of $\gamma$ near $T_R$ and use a temperature independent
value for $\gamma$.

We are concerned here with facets of a vicinal surface $(1,0,p)$,
which correspond to an arrangement of ``primary'' $(1,0,0)$ steps
commensurate with the {\em in plane} periodicity of the underlying
crystal.
(If the crystal was periodic in one direction only, it would be 
a stack of structurless  plates, and
 there would be no $(1,0,p)$ facet.)
This problem is discussed in the accompanying paper \cite{Nozieres01}.
 The use of the universal formula \refe{eq:univ} needs a little care:

\noindent({\em i}) 
$a$ becomes $a(p)=a/p$.
Indeed, for a cubic crystal, one goes from one (1,0,p) plane to the next 
by translating primary steps
by a lattice spacing $a$.

\noindent({\em ii}) 
$\gamma$ is anisotropic and it should be replaced by the 
``surface stiffness''
\begin{equation}
	\gamma_{ij}
	\:=\:
	\gamma\delta_{ij} + \frac{
	\partial^2\gamma}{\partial\theta_i\partial\theta_j}
\end{equation}
After diagonalization of $\gamma_{ij}$, Eq. \refe{eq:univ} is replaced  by
\begin{equation}
	k_BT_R(p) 
	= 
	\frac{2}{\pi}\, \sqrt{\gamma _{x}\gamma _{y}}\, a(p)^2 \: .
\end{equation}
The quantity
$\gamma_x$ is controlled by step compression and it is given by:
\begin{equation}
	\gamma_x\: 
	= 
	\:\frac{1}{a^2}\frac{d^2U}{dn^2}
\end{equation}
As for $\gamma_y$, it appears in the energy cost for an undulation of
steps which move in the $x$ direction by an amount $u=u_0exp(ik_yy)$.
Whenever a step sweeps by a given point the interface is raised by a,
from which we infer a height change $\delta z = na_o\,u$.
If each step rotates by an angle $\zeta = \partial u/\partial y$,
there is both {\it lengthening} and {\it compression} of the steps.
Since one has to replace the energy element $\beta(n)dx$ by
$\beta(n/cos\zeta).(dx/cos\zeta)$, the energy change is of order
$\zeta^2$, and writes
\begin{equation}
	\delta U
	\:=\:
	\frac{\zeta^2}{2} \left(n\beta + n^2\frac{d\beta}{dn}\right)
	\:=\:
	\frac{n\zeta^2}{2}
	\frac{dU}{dn}\:
	=
	\:\frac{1}{2}\gamma_yk_y^2\delta z^2
\end{equation}
from which we infer
\begin{equation}
	\gamma_y\:=\\\frac{1}{n\,a^2}\frac{dU}{dn}
	\,.
\end{equation}
Alltogether, we find
\begin{equation}
	\sqrt{\gamma_x\gamma_y} \,a^2 
	\:=\:
	\sqrt{\frac{U'U''}{n}}
\end{equation}
Since the interaction at small angle, {\em i.e.} small $n$, is
proportional to $1/d^2$, the asymptotic form of $U(n\rightarrow 0)$ is
$n\beta + \delta n^3$ and the effective stiffness writes
$\sqrt{6\beta\delta}/a^2$ in this limit. It is independent of the
tilt angle or step density. It only involves the product of the step
density by the step-step interaction coefficient $\delta$ that can
be inferred from \refe{betaexpansion}.

\begin{figure}[thb]
\begin{center}
\psfig{file=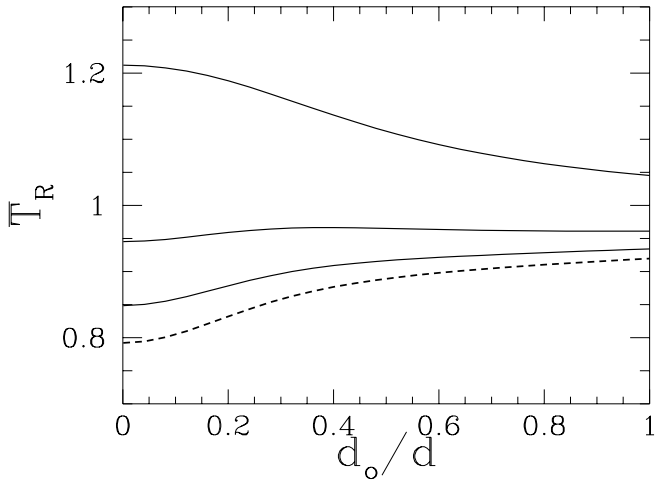,width=9cm,angle=-90}
\end{center}
\caption{
$\overline T_R$ for different values of the parameter $d_o/r_o$ as a function
of $\overline n=d_o/d$. The parameter $d_o/r_o$ is 4, 10, 100.
starting from the bottom. The dashed line corresponds to 
a $d_o/r_o=4$ which is a reasonable value for the experiment
of Ref.~\protect\cite{Pieranski}.
}
\label{fig:TRd}
\end{figure}

We are now equipped to infer $T_R$ from the preceding section. Note
that we do not need to expand in powers of $n$, we can calculate $T_R$
for arbitrary values of $n\,d_0$, the only assumption being $d \gg a$
as before. The $d$ dependence of the factor $\sqrt{U'U''/n}$ is not 
strong and can be conveniently studied by defining the dimensionless 
$\overline T_R$:
\beq
	T_R = 
	{2 \over \pi} \gamma a(p)^2 
	\overline T_R 
	\label{eq:TR}
	\,.
\eeq
The $d\rightarrow \infty$ limit of $T_R$ involves only the 
step energy and the step-step interaction:
\beq
	\overline T_R (d\rightarrow \infty) = 
	{1\over 2} 
	\sqrt{\ln\left( {2 d_o \over  r_o }\right)
	 +\Gamma}	
\eeq
It thus depends on the core radius $r_o$. 
The next order in the development in $1/d$ is quadratic 
and it can change sign as a function of $d_o/r_o$ 
due to the fact that the  $1/d^4$ term
in the interaction energy of two steps is negative [cf. Eq.~\refe{beta2}]. 
In the limit where $d\rightarrow 0$, $\overline T_R=1$. 
The full expression coming from \refe{betalattice}
interpolates between these two limits and it is shown in Figure 
\ref{fig:TRd} for different values of the ratio $r_o/d_o$.
It indicates the departure from a simple
quadratic variation of $T_R$ with the step density $n=1/d$.

The crucial result is that $T_R$ is of the order $\gamma\, a^2$,
despite the fact that elasticity played a central role. 
The lower energy scale $\mu a^3$ is compensated by the lower density 
scale $1/d_o$: since the effective  stiffness has dimension 
$U/n$, elasticity drops out. That might be seen as trivial! 
But a naive argument bared in the usual TLK model would predict 
a step energy $\beta \sim \gamma\, a$ and an elastic 
interaction between steps $\delta\sim (\gamma \,a)^2/\mu d^2$. 
The resulting effective $\gamma$ is then $\gamma(\gamma/\mu\,a )^{1/2}$:
an exact calculation does reduce the scale of $T_R$. As a consequence of 
the small value of the elastic modulus $\mu$, the 
step-step interaction is large, but in the expression for the roughening 
temperature, it appears multiplied by the step 
energy which is small. We have shown that the product of these two 
quantities is nearly independent of $\mu$.

\section{Comparison with the experimental shape of lyotropic crystals}

  Let us finally consider the number of facets observed by Pieranski et
  al.\cite{Pieranski,EPJ}. These crystals show  6 beautiful staircases 
of facets around
  large (1,1,2) facets which we consider as the main ones (see fig.5 in
  ref.\cite{EPJ}). These crystals are cubic with Ia3d symmetry, which
  explains why the main facets are not (1,0,0). The last
facet, i.e. the one with the highest Miller indices, is (9,8,15) (see
fig.7 in ref.\cite{EPJ}) at 293K, the temperature of the experiment. It corresponds to the smallest
tilt angle with respect to the (1,1,2) facet, namely 4 degrees. Pieranski
et al. explain that the
distance between equivalent planes in a [h,k,l] direction is given
by
\begin{equation}
	d_{hkl} 
	\:=\:
	\frac{118}{\sqrt{h^2+k^2+l^2}}\:\mathrm \AA
\end{equation}
This means that the ``primary'' step height is  $a = 48.2 \mathrm \AA$ on
(1,1,2) facets.
As for the ``secondary'' steps on vicinal facets, one finds respective heights
of 7.02, 6.13, 5.44 and 4.89 $\mathrm \AA$ for (8,7,13), (9,8,15),
(10,9,17) and (11,10,19) facets.

The large value of $a$ is responsible
for the large number of facets. Indeed, according to Eq. \refe{eq:univ}, the
roughening temperature of (1,1,2) facets is very high, 26800 K. 
We now understand that room temperature (293K) is very cold compared
to the energy scale involved in the roughening of the main (1,1,2)
facets. If we ignored the correction $\overline{T}_R$ in
Eq. \refe{eq:TR}, we would predict that the last facet at 293K
corresponds to a step height slightly larger than
$a/\sqrt{26800/293}=5.04 \mathrm \AA$. This would allow (10,9,17)
facets to exist at 293 K.

In the case of their crystals, we find a length $d_0 = 3.46 a = 167
\mathrm \AA$. According to our calculation (see fig.\ref{fig:TRd}) the
correction $\overline{T}_R$ is about 0.84 for tilt angles of order 4
degrees, i.e. $d_0/d = 0.24$, so that we predict that the last facet
should have a step height larger than a minimum value 5.50 $\mathrm
\AA$. This is just enough to avoid the existence of (10,9,17) facets,
and allow that of (9,8,15) facets, as seen by Pieranski et
al. Equivalently, we could say that our predictions for the roughening
temperatures are 364K for (9,8,15) surfaces which have thus to be
facetted at 293K, but 287K for (10,9,17) surfaces which should thus be
rough at 293K.
 
Given the number of approximations in our theory, the
agreement with the experiment is surprisingly good. The calculation is
made for a simple cubic lattice where the Burgers vectors are normal to
the main (001) faces. Since Pieranski {\em et al.} 
studied lyotropic crystals  which
have a $Ia3d$ symmetry and tilted Burgers vectors, our model is clearly
oversimplified. We believe that two things should remain if one introduced
such details in a more accurate calculation:\\
1- The highest order facet has to correspond to a step height of order one
tenth of the main one because $T_{R0}$ is about 100 times 293 K.\\
2- The correction $\overline{T}_R$ being smaller than one, one observes
a slightly smaller number of facets than what a simple quadratic law would
predict.
Eventually, it is interesting to explain why we did not consider entropic 
interactions between steps. The origin of this other type of interaction 
is the no-crossing condition of steps\cite{Beg-Rohu}. Its amplitude has 
been calculated\cite{Akutsu} and measured\cite{Rolley}. It is 
\beq
\epsilon^{ent} = \frac{\pi^2}{6} \frac{(k_B T)^2}{\beta} \frac{1}{d^2}
\eeq
Given the low value of the step energy that we have found, one could think 
that the entropic interaction is a significant effect. In the case of soft 
crystals, the ratio of the entropic interaction to the elastic interaction 
is independent of the shear modulus $\mu$. It would be large near the first 
roughening temperature, but it is very small at temperatures two orders of 
magnitude lower.

We are very grateful to Pawel Pieranski and Paul Sotta, who introduced us
to the physics of their crystals and communicated their last results
before publication.

\appendix
\section{Solution of the 2D elastic problem}

In this appendix we provide details on the solution of the two dimensional
elastic problem (every quantity depends only on the depth $z$ and the
coordinate $x)$. We consider a half space isotropic elastic medium $z<0.$
The basic ingredient is the response function relating the displacement $%
u_{i}\left( x\right) $ at the free surface to the force $F_{j}\left(
x^{\prime }\right) $ applied elsewhere on the same surface
\beq
	u_{i}\left( x\right) 
	=
	\int_{-\infty }^{+\infty }\chi _{ij}
	\left( x-x^{\prime}\right) \,F_{j}
	\left( x^{\prime }\right) \,dx^{\prime }
\eeq
The problem is solved via Airy functions (see ref.\cite{Landau})
\beq
	\sigma _{xx}
	=
	\frac{\partial ^{2}A}{\partial z^{2}}\,,\,\sigma _{zz}
	=
	\frac{\partial ^{2}A}{\partial x^{2}}\,,\,\sigma _{xz}
	=
	-\frac{\partial ^{2}A}{\partial x\partial z}
\eeq
For a single Fourier component the biharmonic function $A\left( x,z\right)$
is
\beq
	A\left( x,z\right) 
	=
	\left( a+bz\right) \exp \left[ \left| k\right| z+ikx \right]
\eeq
The constants $a$ and $b$ are fixed by the surface conditions $\sigma
_{zi}\left( 0\right) =F_{i},$ yielding
\beq
	a
	=
	-F_{z}/k^{2} 
	\quad \quad ,\quad \quad 
	b
	=
	i \, F_{x}/k+F_{z}/\left| k \right|
\eeq
From $\sigma _{ij}$ we infer the strain $u_{ij,}$ then the displacement $
u_{x}$ integrating $u_{xx}$.
 We thus know $\partial u_{x}$/$\partial z,$
from which we infer $\partial u_{z}$/$\partial x:$ $u_{z}$ follow upon
integration over $x.$ The final result in momentum space is
\beq
	u_{i}\left( k\right) 
	= 
	\chi _{ij}
	\left( k\right) F_{j}\left( k\right)
\eeq
\beq
	\chi _{zz}
	=
	\chi _{xx}
	=
	\frac{1-\sigma }{\left| k\right| \mu }
	\quad \quad,\quad \quad 
	\chi _{xz}
	=
	\frac{1-2\sigma }{2ik\mu }
\eeq
That same result could be obtained from the response to a point force
given in Ref.\cite{Landau}, integrating first on the dummy coordinate
$y$ and then Fourier transforming.

The displacement and stress fields due to an edge dislocation along the $y$
axis in an infinite medium are well known
\begin{eqnarray*}
	u_{z}^{(o)} 
	&=&
	\frac{b}{2\pi }\left[ \arctan \frac{x}{\overline{z}}+
	\frac{1}{2\left( 1-\sigma \right) }
	\frac{x\overline{z}}{x^{2}+\overline{z}^{2}}
	\right]  
	\\
	u_{x}^{(o)} 
	&=& 
	-\frac{b}{2\pi }\left[ 
	\frac{1-2\sigma }{4\left( 1-\sigma \right) }
	\log \left( x^{2}+\overline{z}^{2}\right) 
	+\frac{1}{2\left(1-\sigma \right) }
	\frac{\overline{z}^{2}}{x^{2}+\overline{z}^{2}}\right]
	\\
	\sigma _{zz}^{(o)} 
	&=&
	-\,\frac{\mu b}{2\pi \left( 1-\sigma \right) }~
	\frac{x\left( 3\overline{z}^{2}+x^{2}\right) }{
	\left( x^{2}+\overline{z}^{2}\right) ^{2}}
	\\
	\sigma _{xz}
	&=&
	\frac{\mu b}{2\pi 
	\left( 1-\sigma \right) }~\frac{\overline{z}
	\left( \overline{z}^{2}-x^{2}\right) }
	{\left(x^{2}+\overline{z}^{2}\right) ^{2}}
\end{eqnarray*}
$\overline{z}$ is the distance from the core. If the dislocation is buried
at a depth $h$ below the origin, $\overline{z}=z+h.$ The corresponding
stress and displacement at $z=0$ are in Fourier space
\begin{eqnarray*}
\sigma _{zx}^{(o)}\left( k\right)  &=&\frac{\mu b}{4\pi \left( 1-\sigma
\right) }\left| k\right| h\exp \left[ -\left| k\right| h\right]  \\
\sigma _{zz}^{(o)}\left( k\right)  &=&i\frac{\mu b\text{sgn}\left( k\right)
}{4\pi \left( 1-\sigma \right) }\left[ 1+\left| k\right| h\right] \exp
\left[ -\left| k\right| h\right]  \\
u_{z}^{(o)}\left( k\right)  &=&-\frac{ib}{4\pi k}\left[ 1+\frac{\left|
k\right| h}{2\left( 1-\sigma \right) }\right] \exp \left[ -\left| k\right|
h\right]
\end{eqnarray*}
From that we infer the extra displacement $u_{z}^{(1)}\left( k\right) $ due
to relaxation of counterforces, in which we identify the contributions due
to $F_{z}^{(o)}$ and to $F_{x}^{(o)}$%
\begin{eqnarray*}
	u_{z}^{(1x)}	
	\left( k\right)  
	&=&
	-\chi _{zx}\left( k\right) 
	\sigma_{zx}^{(o)}\left( k\right) 
	=
	\frac{b\left( 1-2\sigma \right) }{8\pi ik
	\left(1-\sigma \right) }
	\left| k\right| h e^{-|k|h}  
	\\
	u_{z}^{(1z)}\left( k\right)  
	&=&
	-\chi _{zz}\left( k\right) 
	\sigma_{zz}^{(o)}\left( k\right) 
	=
	-\,\frac{ib}{4\pi k}
	\left[ 1+\left| k\right|h\right] 
	e^{-|k|h}  	
\end{eqnarray*}
The net surface profile as controlled by elasticity is
\beq
u_{z}^{\rm el}=u_{z}^{(o)}+u_{z}^{(1x)}+u_{z}^{(1z)}
\eeq
It is easily verified that $u_{z}^{\rm el}=2u_{z}^{(1z)}:$ the shape of the
surface is controlled by by $F_{z}^{(o)}$ only. Such a very simple result is
not an accident: instead of using a single dislocation with two
counterforces $F_{z}^{(o)}$ and $F_{x}^{(o)},$ we could have used an image
picture with a doubled tangential counterforce $2F_{x}^{(o)}.$ The
displacement would then read $u_{z}^{\rm el}=2u_{z}^{(o)}+2u_{z}^{(1x)}$;
identification of the two expressions of $u_{z}^{\rm el}$ yields the above
result! (As mentioned in the text explicit comparison of the approaches at
every stage of the calculation is an useful trick for hunting sign mistakes).

In addition to the surface profile, we need the net elastic energy which we
write as
\begin{eqnarray*}
	\lefteqn{
	U^{\rm el} 
	=
	U^{(o)}
	}\nonumber \\
	&-&\,2\pi \int_{-\infty }^{+\infty }dk\,\,
	\left[ \left| \chi_{zz} 
	\left( k\right) \sigma _{zz}^{(o)}
	\left( k\right) \right| ^{2}+\chi_{xx}
	\left( k\right) \left| \sigma _{zx}^{(o)}
	\left( k\right) \right|^{2}\right]  
	\\
	&=&
	\frac{\mu a^{2}}{4\pi \left( 1-\sigma \right) }
	\int_{0}^{1/r_o}\frac{dk}{k}
	\left\{ 1-\left[ 1+2kh+2k^{2}h^{2}\right] 
	e^{-2kh}
	\right\}
\end{eqnarray*}
The infrared divergence at $k=0$ cancels out (the upper distance cut off is $%
h$ instead of $R_{\max }),$ but the large $k$ divergence remains: it
describes the dislocation core! The last step is inclusion of the capillary
correction, as described in the main text. The relevant quantity is
\beq
\alpha _{k}=\gamma k^{2}\chi _{zz}\left( k\right) =kd_{o}
\eeq
Since the relevant $k$ are $\approx 1/d_{o},$ the need for a selfconsistent
calculation is apparent. The net capillary energy, as ''screened'' by
elastic response, is
\begin{eqnarray*}
	U^{\rm cap} 
	&=&
	2\pi \int_{-\infty }^{+\infty }\!\!\!\!\!\!dk
	\left\{ \frac{\gamma k^{2}}{2}%
	\left| u_{z}^{\rm el}\left( k\right) +u_{z}^{\rm cap}
	\left( k\right) \right| ^{2}+%
	\frac{\left| u_{z}^{\rm cap}\left( k\right) \right| ^{2}}%
	{2\chi _{zz}\left(k\right) }\right\}  \\
	&=&
	2\pi \int_{-\infty }^{+\infty }\!\!\!\!\!\!dk
	\frac{\gamma k^{2}}{2\left( 1+\alpha_{k}\right) }
	\left| u_{z}^{\rm el}(k)\right| ^{2}
\end{eqnarray*}
The expression for the total energy used in the text follows at once.


\begin{thebibliography}{}
\bibitem{remark} 
Our present definition of ``soft crystals'' is based
on a comparison of elasticity to surface tension. Helium crystals are
not soft in this sense, because their interface with liquid helium has
a small energy \cite{Balibar94}.
\end{thebibliography}
\end{document}